%% file: main.tex
\documentclass[runningheads]{llncs}
\usepackage{multirow}
\usepackage[usenames,dvipsnames]{color}
\usepackage[table]{xcolor}
\usepackage[final]{graphicx}
\usepackage{amssymb}
\usepackage{pifont}
\usepackage{array}
 \usepackage{wrapfig}
 \usepackage{paralist}
 \usepackage{cancel}
\begin{document}

\title{Same Stats, Different Graphs\\ {\normalsize (Graph Statistics and Why We Need Graph Drawings)}}

\titlerunning{Same Stats, Different Graphs}  
%
\author{Hang Chen$^1$, Utkarsh Soni$^2$, Yafeng Lu$^2$, Ross Maciejewski$^2$, \\Stephen Kobourov$^1$}
\authorrunning{Chen {\em et al.}} 
%
\tocauthor{}
\institute{$^1$University of Arizona, Tucson, AZ, USA \\kobourov@cs.arizona.edu\\
$^2$Arizona State University, Tempe, AZ, USA \\
}

\maketitle              

\vspace{-.15cm}\begin{abstract}
Data analysts commonly utilize statistics to summarize large datasets. While it is often sufficient to explore only the summary statistics of a dataset (e.g., min/mean/max), Anscombe's Quartet demonstrates how such statistics can be misleading. We consider a similar problem in the context of graph mining.
To study the relationships between different graph properties and statistics, we examine all 
low-order ($\leq 10$)
non-isomorphic graphs and provide a simple visual analytics system to explore correlations across multiple graph properties. However, for graphs with more than ten nodes, generating the entire space of graphs becomes quickly intractable. We use different random graph generation methods to further look into the distribution of graph statistics for higher order graphs and investigate the impact of various sampling methodologies.
We also describe a method for generating many graphs that are identical over a number of graph properties and statistics yet are clearly different and identifiably distinct.

\end{abstract}

\vspace{-.4cm}
\section{Introduction}
\vspace{-.2cm}
\begin{wrapfigure}[13]{r}{0.5\textwidth} \vspace{-1.6cm}\includegraphics[width=.52\textwidth]{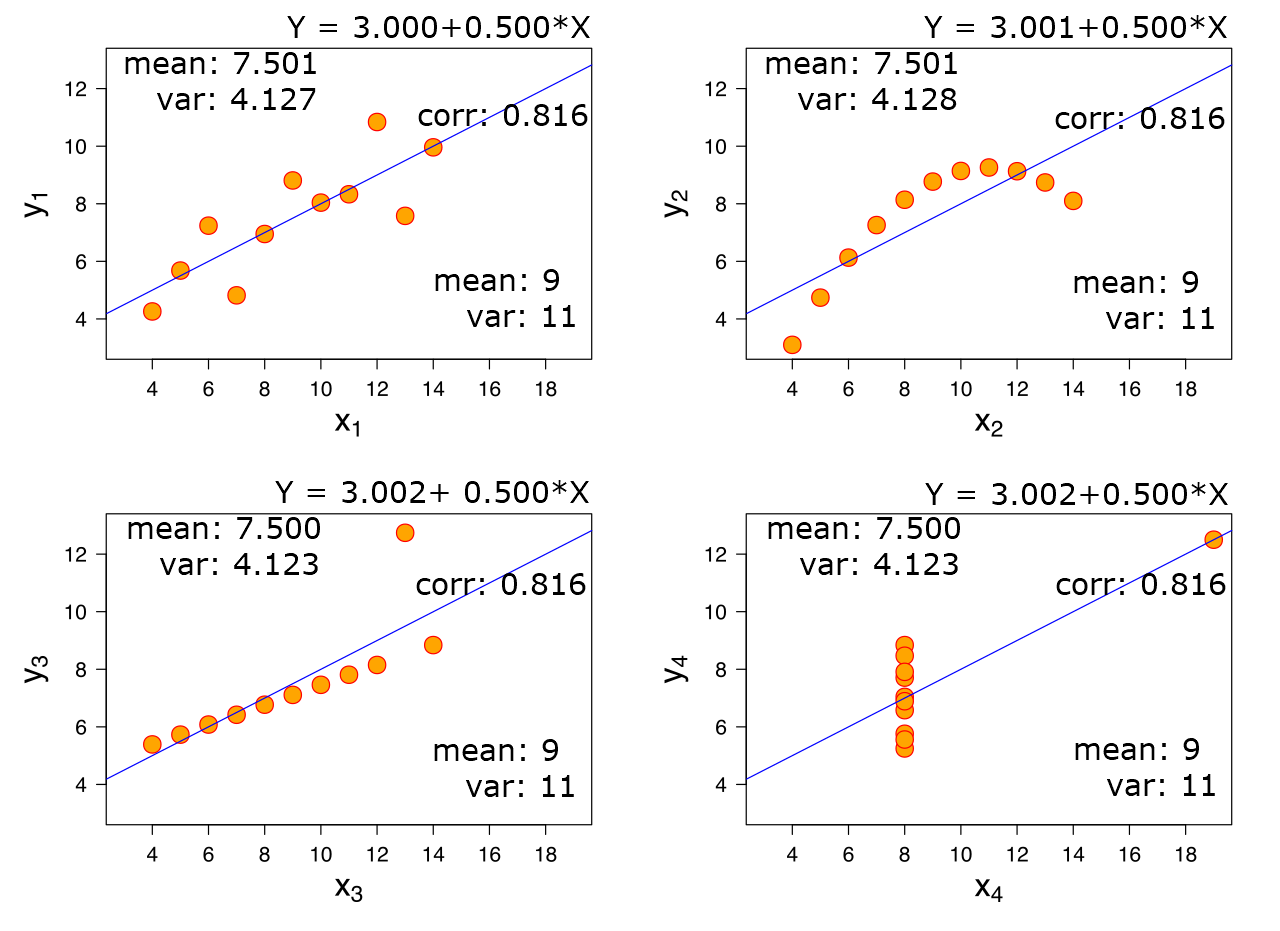}
\vspace{-.65cm}\caption{Anscombe's quartet: All four datasets have the same mean and st.~deviation in $x$ and $y$ and $(x,y)$-correlation.\label{fig:AC}}
\centering
\end{wrapfigure}Statistics are often used to summarize a large dataset. In a way, one hopes to find the ``most important" statistics that capture one's data. For example, when comparing two countries, we often specify the population size, GDP, employment rate, etc. The idea is that if two countries have a ``similar" statistical profile, they are similar (e.g., France and Germany have a more similar demographic profile than France and USA). However, Anscombe's quartet~\cite{Anscombe}  convincingly illustrates that datasets with the same values over a limited number of statistical properties can be fundamentally different -- a great argument for the need to visualize the underlying data; see Fig.~\ref{fig:AC}.

\begin{wrapfigure}[28]{r}{0.61\textwidth} \vspace{-.0cm}
\fbox{\includegraphics[width=0.28\textwidth]{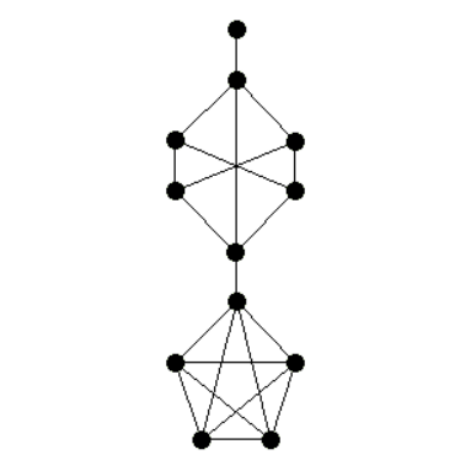}}\fbox{\includegraphics[width=0.28\textwidth]{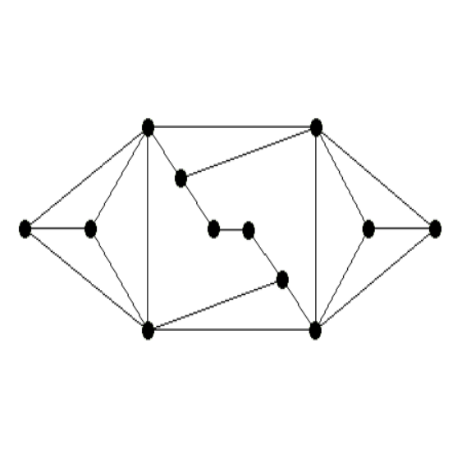}}
\fbox{\includegraphics[width=0.28\textwidth]{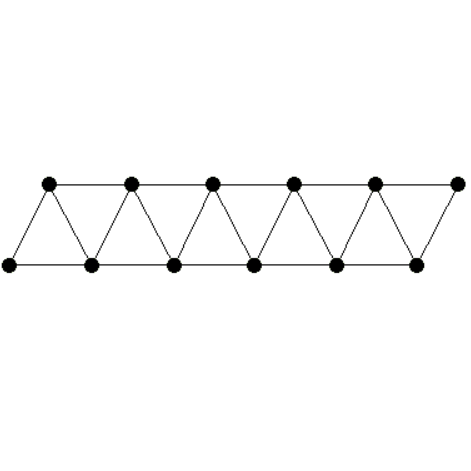}}\fbox{\includegraphics[width=0.28\textwidth]{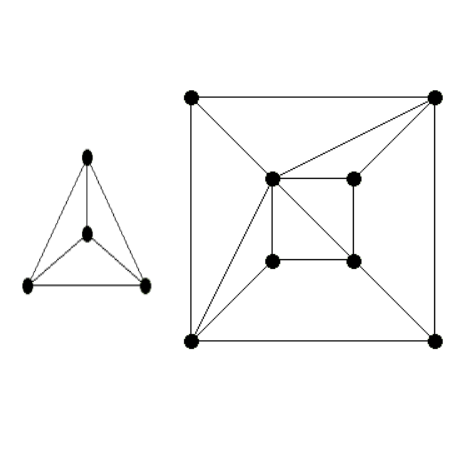}}
  \caption{These four graphs share the same 5 common statistics: $|V| = 12$, $|E| = 21$, number of triangles $|\bigtriangleup|=10$, girth $=3$ and global clustering coefficient {\em GCC}$=0.5$.
 However, structurally the graphs are very different: some are planar others are not, some show regular patterns and are symmetric others are not, and finally, one of the graphs is disconnected, another is 1-connected and the rest are 2-connected.\label{fig:HC}}
\end{wrapfigure}
Similarly, in the graph analytics community, a variety of statistics are being used to summarize graphs, such as graph density, average path length, global clustering coefficient, etc. However, summarizing a graph with a fixed set of graph statistics leads to the problem illustrated by Anscombe. 
It is easy to construct several graphs that have the same basic statistics (e.g., number of vertices, number of edges, number of triangles, girth, clustering coefficient) while the underlying graphs are clearly different and identifiably distinct; see Fig.~\ref{fig:HC}. 
From a graph theoretical point of view, these graphs are very different: they differ in  connectivity, planarity, symmetry, and other structural properties.

Recently, Matejka and Fitzmaurice~\cite{matejka2017same} proposed a dataset generation method that can modify a given 2-dimensional point set (like the ones in Anscombe's quartet) while preserving its summary statistics but significantly changing its visualization (what they call ``graph"). Given the graphs in Fig.~\ref{fig:HC}, we consider whether it is also possible to modify a given graph and preserve a given set of summary statistics while significantly changing other graph properties and statistics.
Note that the problem is much easier for 2D point sets and basic statistics, such as mean, deviation and correlation, than for graphs where many graph properties are structurally correlated (e.g., diameter and average path length).
With this in mind, we first consider how can we fix a few graph statistics (such as number of nodes, number of edges, number of triangles) and vary another statistic (such as clustering coefficient or connectivity). We find that there is a spectrum of possibilities. Sometimes the ``unrestricted" statistic can vary dramatically, sometimes not, and the outcome depends on two issues: (1) the inherent correlation between some statistics (e.g., density and number of triangles), and (2) the bias in graph generators.

We begin by studying the correlation between graph summary statistics across the set of all non-isomorphic graphs with up to 10 vertices. 
The statistical properties derived for all graphs for a fixed number of vertices provide further information about certain ``restrictions." In other words, the range of one statistic may be restricted if another statistical property is fixed. However, we cannot explore the entire space of graph statistics and correlations. As the number of vertices grows, the number of different non-isomorphic graphs grows super-exponentially. For $|V|=1, 2 \dots 9$ the numbers are $1, 2, 4, 11, 34, 156, 1044, 12346,$ $274668$, but already for $|V|=16$ we have $6\times 10^{22}$ 
non-isomorphic graphs. 

To go beyond ten vertices we use graph generators based on 
models, such as Erd\"os-R\'enyi and Watts-Strogatz. However, different graph generators have different biases and these can  significantly impact the results. 
We study the extent to which sampling using random generators can represent the whole graph set for an arbitrary number of vertices with respect to their coverage of the graph statistics. One way to evaluate the performance of random generators is based on the ground-truth graph sets that are available: all non-isomorphic graphs for $|V|\leq 10$ vertices. If we randomly generate a small set of graphs (also for  $|V|\leq 10$ vertices) using a given graph generator, we can explore how well the sample and generator cover the space of graph statistics. In this way, we can begin exploring the issues of ``same stats, different graphs'' for larger graphs.

 Data and tools are available at \url{http://vader.lab.asu.edu/sameStatDiffGraph/}.
Specifically, we have a basic visual analytics system and basic exploration tools for the space of all low-order ($\leq 10$)
non-isomorphic graphs and sampled higher order
graphs. We also include a generator for ``same stats, different graphs," i.e., multiple graphs that are identical over a number of graph statistics, yet are clearly different.

\vspace{-.4cm}
\section{Related Work}
\vspace{-.2cm}
We briefly review the graph mining literature, paying special attention to commonly collected graph statistics. We also consider different graph generators.

\input{input/propertytable}

\smallskip\noindent{\bf Graph Statistics:}  
Graph mining is applied in different domains from bioinformatics and chemistry, to software engineering and social science. Essential to graph mining is the efficient calculation of various graph properties and statistics that can provide useful insight about the structural properties of a graph. A review of recent graph mining systems  identified some of the most frequently extracted statistics. We list those, along with their definitions, in Table~\ref{Table:properties}. These properties range from basic, e.g., vertex count and  edge count, to complex, e.g., clustering coefficients and average path length. Many of them can be used to derive further properties and statistics. 
For example, graph density can be determined directly as the ratio of the number of edges $|E|$ to the maximum number of edges possible 
$|V|\times (|V| - 1)/2$, 
and real-world networks are often found to have a low graph density~\cite{melancon2006just}. Node connectivity and edge connectivity measures may be used to describe the resilience of a network~\cite{cartwright1956structural,loguinov2003graph}, 
and graph diameter~\cite{hanneman2005introduction} captures the maximum among all pairs of shortest paths~\cite{albert1999internet,broder2000graph}. 

Other graph statistics measure how tightly nodes are grouped in a graph. For example,  clustering coefficients have been used to describe many real-world networks, and can be measured locally and globally. Nodes in a highly connected clique tend to have a high local clustering coefficient, and a graph with clear clustering patterns will have a high global clustering coefficient~\cite{feld1981focused,frank1982cluster,karlberg1997testing,newman2003structure}.
Studies have shown that the global clustering coefficient
has been found to be nearly always larger in real-world graphs than in Erd\"os-R\'enyi graphs with the same number of vertices and edges~\cite{chakrabarti2006graph,newman2003structure,uzzi2005collaboration}, and a small-world network should have a relatively large average clustering coefficient~\cite{davis2003small,ebel2002scale,watts1998collective}. 
The average path length (APL) is also of interest; small-world networks have APL that is logarithmic in the number of vertices, while real-world networks have small (often constant) APL~\cite{davis2003small,ebel2002scale,newman2003structure,uzzi2005collaboration,van2004yeast,watts1998collective}. 

Degree distribution is one frequently used property describing the graph degree statistics. Many real-world networks, including communication, citation, biological and social networks, have been found to follow a power-law shaped degree distribution
\cite{boccaletti2006complex,chakrabarti2006graph,newman2003structure}. Other real world networks have been found to follow an exponential degree distribution \cite{guimera2003self,sen2003small,wei2009worldwide}.
Degree assortativity is of particular interest in the study of social networks and is  
calculated based on the Pearson correlation between the vertex degrees of connected pairs~\cite{newman2002assortative}. A random graph generated by Erd\"os-R\'enyi model has an expected assortative coefficient of $0$.   
Newman~\cite{newman2002assortative} extensively studied assortativity in real-world networks and found that social networks are often assortative (positive assortativity), i.e., vertices with a similar degree preferentially connect together, whereas technological and biological networks tend to be disassortative (negative assortativity) implying that vertices with a smaller degree tend to connect to high degree vertices. Assortativity has been shown to affect clustering~\cite{maslov2004detection}, resilience~\cite{newman2002assortative}, and epidemic-spread~\cite{boguna2002epidemic} in networks.


\smallskip\noindent{\bf Graph Generators:} 
Basic graph statistics have been used to describe various classes of graphs (e.g., geometric, small-world, scale-free) and a variety of algorithms have been developed to automatically generate graphs that mimic these various properties.
Charkabati et al.~\cite{chakrabarti2007graph} divide graph models and generators into four broad categories:
\begin{compactenum}
\item Random Graph Models: The graphs are generated by a random process.
\item Preferential Attachment Models: In these models, the ``rich get richer," as the network grows, leading to power law effects. 
\item Optimization-Based Models: Here, power laws are shown to evolve when risks are minimized using limited resources. 
\item Geographical Models: These models consider the effects of geography (i.e., the positions of the nodes) on the topology of the network. This is relevant for modeling router or power grid networks.
\end{compactenum}

The Erd{\"o}s-R\'enyi (ER) network model is a simple graph generation model~\cite{chakrabarti2006graph} that creates graphs either by choosing a network randomly with equal probability from a set of all possible networks of size $|V|$ with $|E|$ edges~\cite{gilbert1959random} or by creating each possible edge of a network with $|V|$ vertices with a given probability $p$~\cite{erdos1959random}. The latter process gives a binomial degree distribution that can be approximated with a Poisson distribution. 
Note that fixing the number of nodes and using $p = 1/2$ results in a good sampling of the space of isomorphic graphs. However, this model (and others discussed below) does not sample well the space of non-isomorphic graphs, which are the subject of our study. 


Watts and Strogatz~\cite{watts1998collective} addressed the low clustering coefficient limitation of the ER model in their model (WS) which can be used to generate small-world graphs.
The WS model can generate disconnected graphs, but the variation suggested by Newman and Watts~\cite{newman1999scaling} ensures connectivity.
Models have also been proposed for generating synthetic scale-free networks with a varying scaling exponent($\gamma$). The first scale-free directed network model was given by de Solla Price~\cite{price1976general}. Barabasi and Albert (BA)~\cite{barabasi1999emergence} described another popular network model for generating undirected networks. It is a network growth model in which each added vertex has a fixed number of edges $|E|$, and the probability of each edge connecting to an existing vertex $v$ is proportional to the degree of $v$. 
Dorogovtsev et al.~\cite{dorogovtsev2000structure} and Albert and Barabasi~\cite{albert2002statistical} also developed a variation of the BA model with a tunable scaling exponent.    

Bach et~al.~\cite{bach2012interactive} introduce an interactive system to create random graphs that match user-specified statistics based on a genetic algorithm. The statistics considered are $|V|, |E|$, average vertex degree, number of components, diameter, ACC, density, and the number of clusters (as defined by Newman and Girvan~\cite{girvan2002community}). The goal is to generate graphs that get as close as possible to a set of target statistics; however, there are no guarantees that the target values can be obtained. 
Somewhat differently, we are interested in creating graphs that match several target statistics exactly, but differ drastically in other parameters. 

\vspace{-0.2cm}
\section{Preliminary Experiments and Findings}
\vspace{-0.2cm}

In a recent study of the ability to perceive different graph properties such as edge density and clustering coefficient in different types of graph layouts (e.g., force-directed, circular), we generated a large number of graphs with 100 vertices. Specifically, we generated graphs that vary in a controlled way in edge density and graphs that vary in a controlled way in the average clustering coefficient~\cite{soni2018perception}. A post-hoc analysis of this data (http://vader.lab.asu.edu/GraphAnalytics/), reveals some interesting patterns among the statistics described in Table~\ref{Table:properties}. 

\begin{figure}[t]
\centering
\includegraphics[width = 0.49\textwidth]{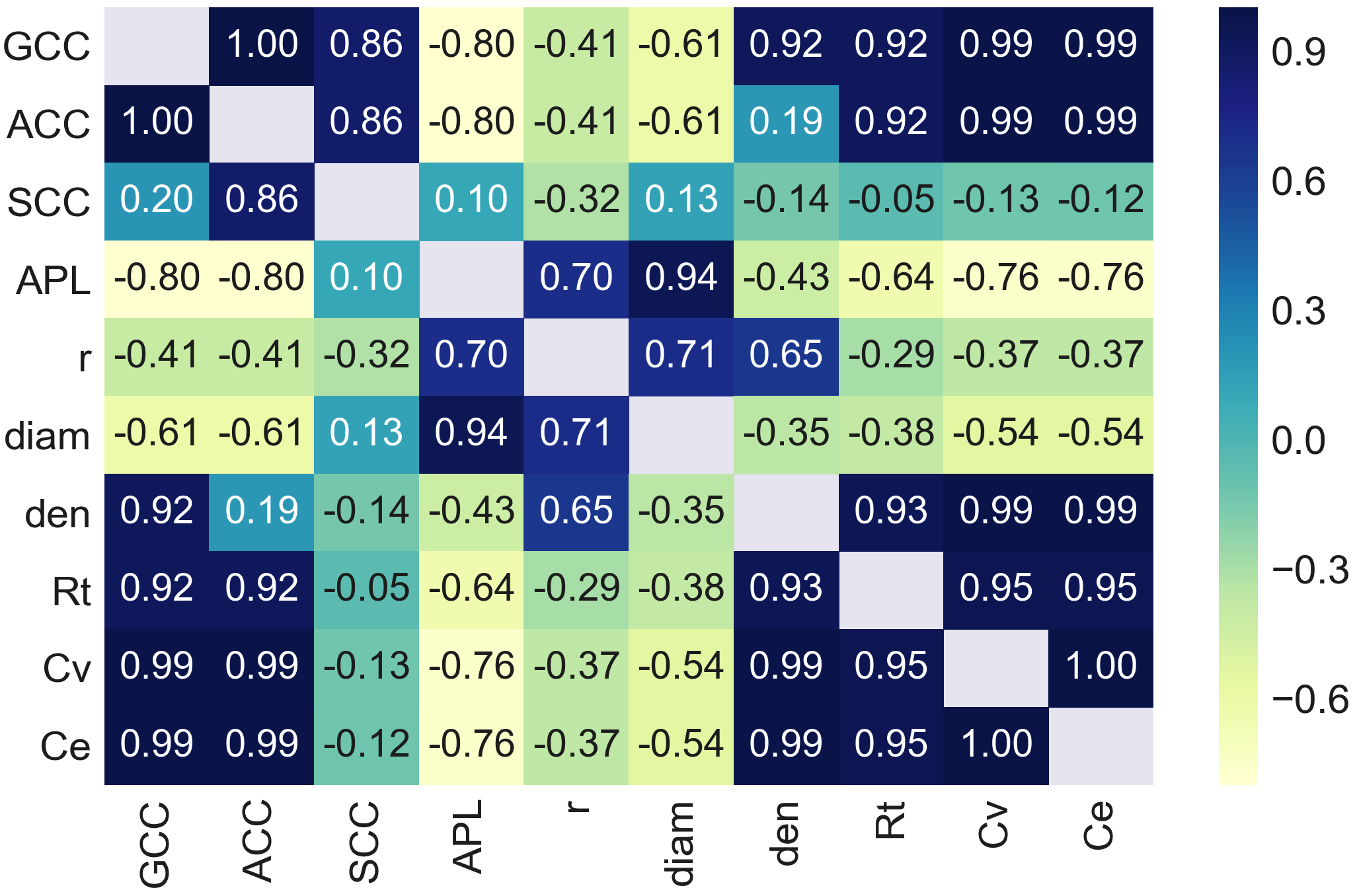}
\hfill
\includegraphics[width = 0.49\textwidth]{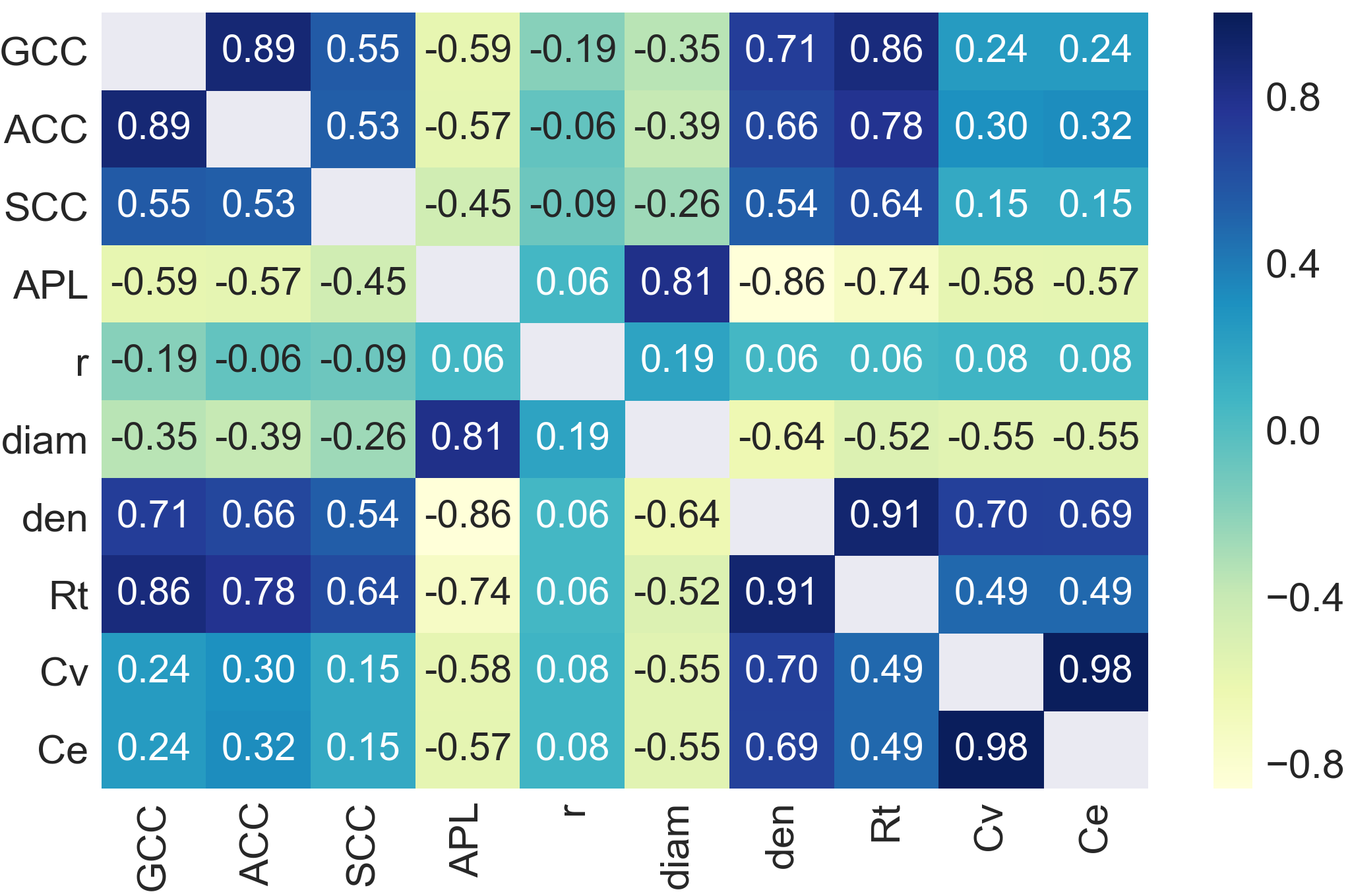}
\caption{Graph property correlation matrix plots for the edge density dataset (left) and the ground truth set of all non-isomorphic graphs on $|V|=9$ vertices (right). 
}
\label{Fig:heatmap}
\end{figure}

The edge density dataset has 4,950 graphs and 
We compute all ten statistics from Table~\ref{Table:properties} and compute Pearson correlation coefficients; see Fig~\ref{Fig:heatmap}.
We observed high positive (blue) correlations and negative (yellow) correlations for many property pairs. 
For example, the average clustering coefficient is highly correlated with the global clustering coefficient, the number of triangles, and graph connectivity.

Note, however, these graphs were created for a very specific purpose and cover only limited space of all graphs with $|V|=100$. 
The type of generators we used, and the way we used them (some statistical properties were controlled), could bias the results and influence the correlations. 
%
The fact that these correlations exist when some properties are fixed indicates that we can keep certain graph statistics fixed while manipulating others. This motivated us to conduct the following experiments:
\begin{enumerate}
\item Generate all non-isomorphic lower order graphs ($|V|\leq 10$) and analyze the relationships between statistical properties. We consider this type of data as ground truth due to its completeness.
\item Use different graph generators 
and compare how well they represent the space of non-isomorphic graphs and how well they cover the range of possible values in the ground truth data.
\end{enumerate}


An analysis of the set of 274,668 non-isomorphic graphs on $|V|=9$ vertices shows that the correlations are quite different than those in graphs from our edge density experiment; see Fig.~\ref{Fig:heatmap}.

\begin{figure}[th]
\vspace{-.1cm}
\includegraphics[width=\linewidth]{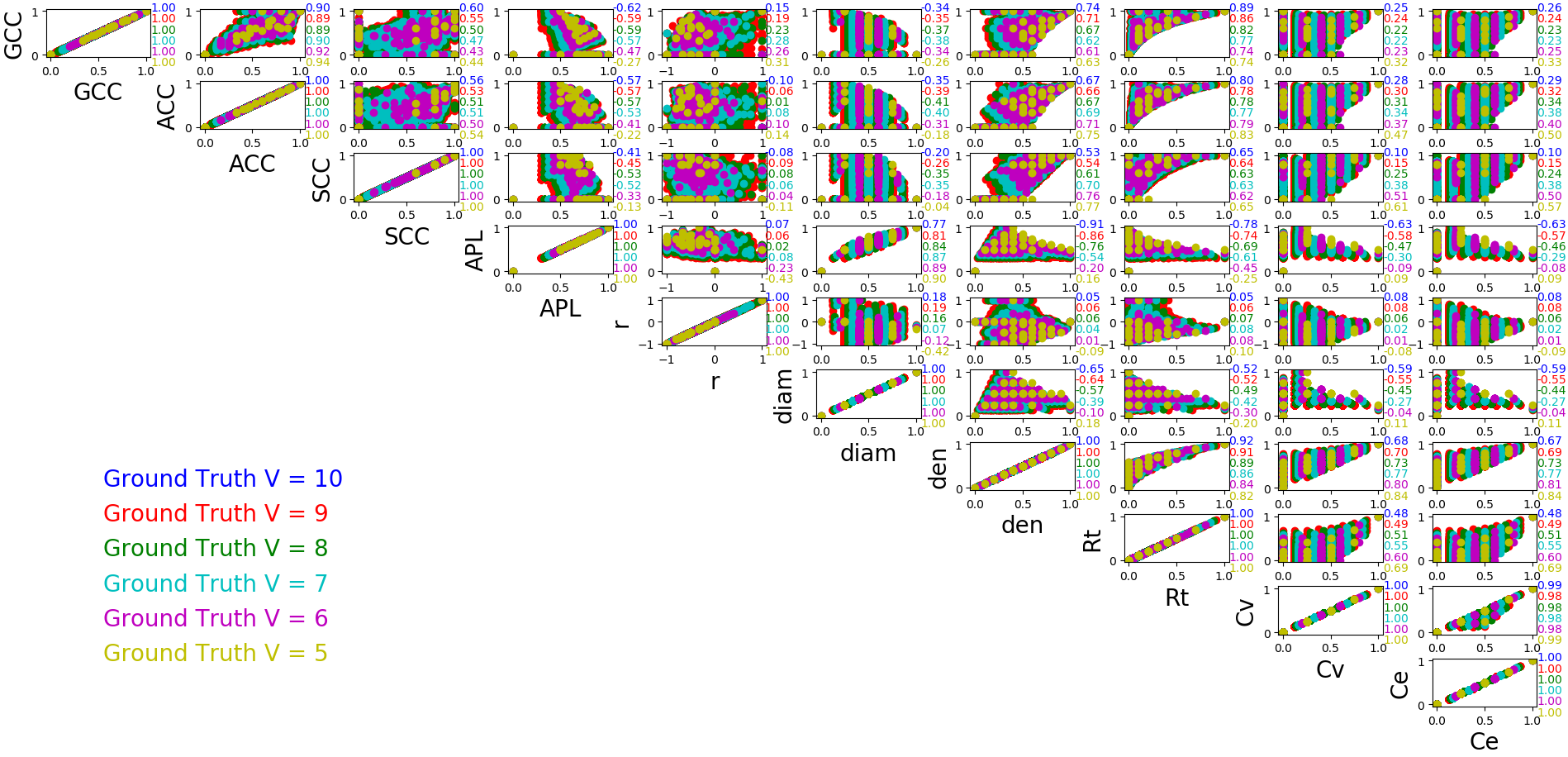}
\caption{Correlations between graph statistics in the ground truth for $|V|=5,6,7,8,9$. Note that for $|V|=9$ there are already 274,668 points. Points are plotted to overlap, with the largest sets plotted first (i.e., $|V|=9, ... |V|=5$) to enable us to identify the range of statistics that can be covered with a given number of vertices.}
  \label{fig:hang-style}
\end{figure}

\section{Analysis of Graph Statistics for Low-Order Graphs}

We start the experiment by looking at pairwise relationship of graph statistics of low-order graphs, where all non-isomorphic graphs can be enumerated. If two statistics, say $s_1$ and $s_2$, are highly correlated, then fixing $s_1$ is likely to restrict the range of possible values for $s_2$. On the other hand, if $s_1$ and $s_2$ are independent, fixing $s_1$ might not impact the range of values for $s_2$, yielding same stats ($s_1$) different graphs ($s_2$). With this in mind, we first study the correlations between the statistics under consideration.

We compute all statistics for all non-isomorphic graphs on $|V|=4,5, \ldots, 10$ vertices (we exclude graphs with fewer vertices as many of the statistics are not well defined and there are only a handful of graphs).
We then consider the pairwise correlations between the different statistics
and how this changes as the graph order increases; 
see Fig~\ref{fig:hang-style}.
To compare the coverage of statistics with different $|V|$, we scale the statistic values into the same range. By definition, clustering coefficients (ACC, GCC, SCC) are in the $[0,1]$ range and degree assortativity is in the $[-1, 1]$ range. We keep their values and ranges without scaling.  
Edge density, number of triangles, diameter and  connectivity measures ($C_v$ and $C_e$), are normalized into $[0,1]$ (dividing by the corresponding maximum value). 
The last statistic, APL, is also normalized into $[0,1]$, subject to some complications:  we compute the exact average path length to divide by in our ground truth datasets, but not when we use the generators, where we use the maximal path length encountered instead (which may not be the same as the maximum).

\begin{wrapfigure}[21]{r}{0.6\textwidth} 
\vspace{-.1cm}\includegraphics[width=0.95\linewidth]{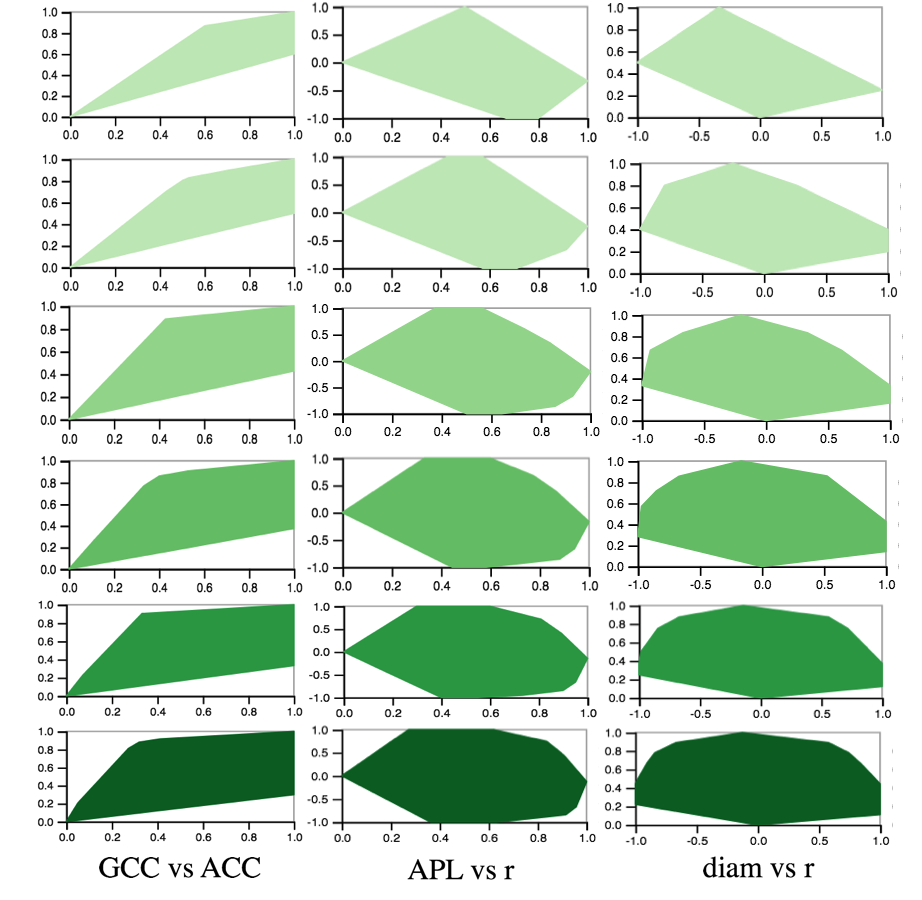}
\vspace{-.5cm}\caption{The convex hull of graph coverage across several statistical properties. Each row (starting from the top) represents all graphs for a fixed number of vertices ($|V| = 5 ... |V|=10$). Columns are pairs of graph properties.\vspace{-.4cm}}
  \label{fig:coverage_comparison}
\end{wrapfigure}

\begin{figure} [!t]
\vspace{-.4cm}
\includegraphics[width=\linewidth]{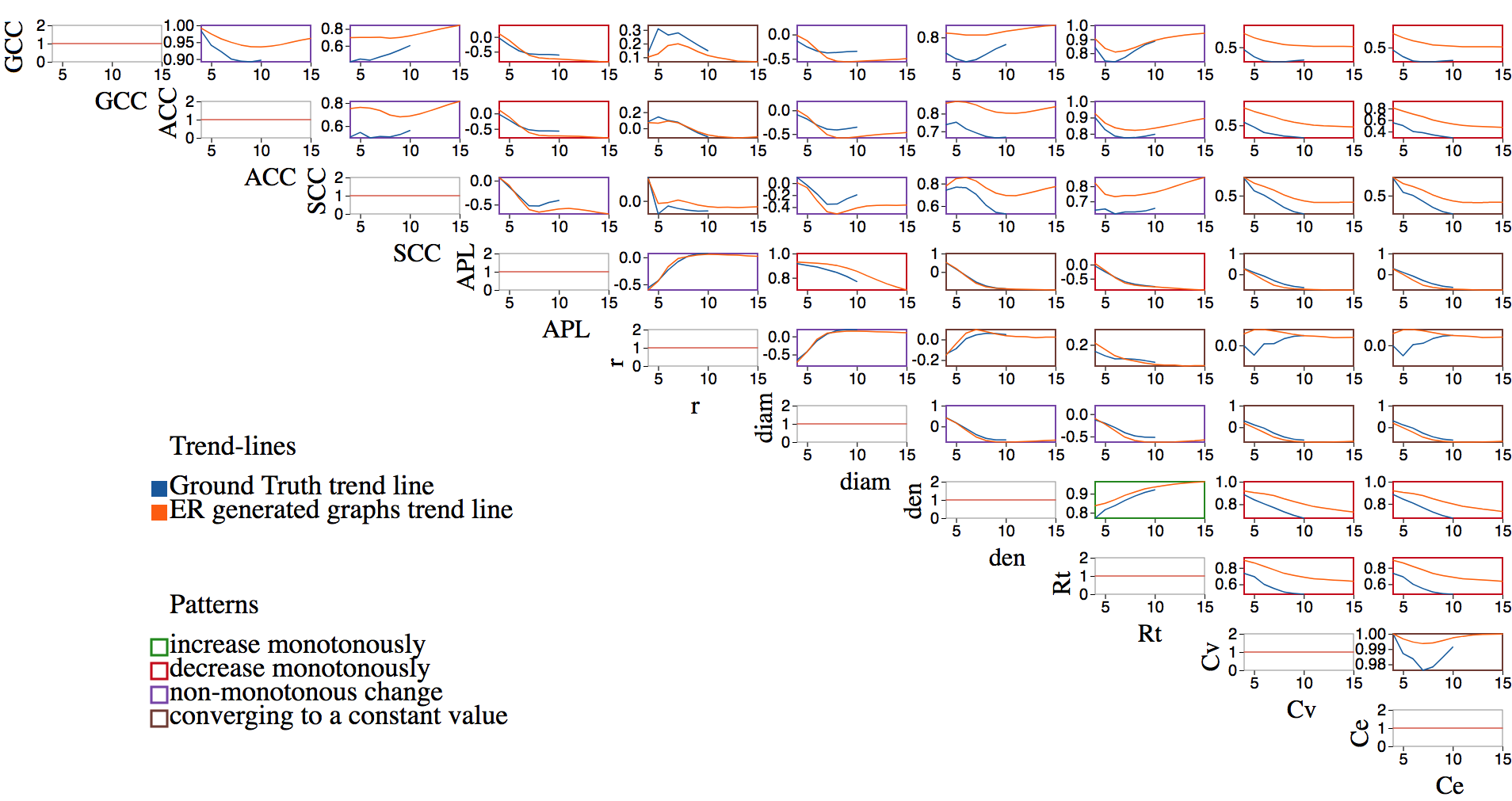}
\vspace{-.7cm}
\caption{Trends in the correlations with increasing $|V|$: the $x$-axis shows the number of vertices and the $y$-axis shows the correlation value for the pair of graph statistics.}
  \label{fig:correlation_vs_n}
\end{figure}

It is easy to see that the coverage of values expands with increasing $|V|$. Figure \ref{fig:coverage_comparison} shows this pattern for three pairs of properties. 
This indicates that we are more likely to find larger ranges of different statistics for graphs with more vertices given the same set of fixed statistics.
With this in mind, we consider graphs with more than 10 vertices, but this time relying on random graph generators. 
Figure~\ref{fig:correlation_vs_n} shows how correlation values between all pairs of statistics change when the number of vertices increases. The blue trend lines for the ground truth data show the correlation values calculated using the set of all possible graphs for a given number of nodes. The orange
trend lines show the correlation values calculated from graphs generated with the ER model. Specifically, the ER data is created as follows: for each value of $|V|=5,6,\ldots, 15$ we generate $100,000$ graphs with $p$ selected uniformly at random in the $[0,1]$ range.

For most of the cells in the matrix shown in Figure \ref{fig:correlation_vs_n}, the correlation values seem to converge 
as $|V|$ becomes larger than $8$. 
(both in the ground truth and the ER-model generated graph sets). 
Moreover, for most  of the cells, the pattern of the change in correlation values appears to be the same for both sets. 
 Analyzing the trend lines of the ER-model, we observe four patterns of change in the correlation values: convergence to a constant value, monotonic decrease, monotonic increase, and non-monotonic change.
These patterns are highlighted in Figure~\ref{fig:correlation_vs_n} by enclosing boxes of different colors.
There are exceptions that do not fit these patterns, e.g., ($S_c$, r) and in two cases, (r, $C_v$) and (r, $C_e$), the trend lines show different patterns.

\vspace{-.4cm}
\section{Graph Statistics and Graph Generators}
\vspace{-.2cm}
While we can explore statistical coverage and correlations in low-order graphs, it is difficult to generate all non-isomorphic graphs with more than 10 vertices due to the super-exponential increase in the number of different graphs (e.g., for $|V|=16$ we there are $6\times 10^{22}$ different graphs). However, these higher order graphs are common in many domains. As such, we want to further explore this issue of ``same stats, different graphs" for larger graphs. As such, we turn to graph generators to help us explore the same-stats-different-graphs problem. 


We select four different random generators that cover 
the four categories~\cite{chakrabarti2007graph} of graph generation:
the ER random graph model,  the WS small-world model, the BA preferential attachment model, and the geometric random graph model.  

\smallskip\noindent{\bf Coverage for Ground-Truth Graph Set:} 
We use implementations of all four generators (ER, WS, BA, geometric) from NetworkX~\cite{hagberg2008exploring}, with three variants of ER ($p=0.5$, $p$ selected uniformly at random from the $[0,1]$ range, and $p$ selected to match edge density in the ground truth). More details about the graph generators and how well they perform for our tasks are  provided in the full version of the paper~\cite{2018arXiv180809913C}.
 For each generator, we generate 1\%, 0.1\% and 0.01\% of the total number of graphs in ground-truth graph set. We use low sampling rates as for high order graphs the ground truth set is huge and any sampling strategy will have just a fraction of the total. 
 Our goal here is to explore whether a small sample of graphs could be representative of the ground truth set of non-isomorphic graphs and cover the space well.
 
We evaluate the different graph generators in two different ways. First we want to see whether a graph generator is {\em representative} of the ground truth data, i.e., whether the generator yields a sample that with similar properties as those in the ground truth. Second, we want to see whether a graph generator is {\em covering} the complete range of values found in the ground truth data. 

\begin{figure}
\includegraphics[width = \textwidth]{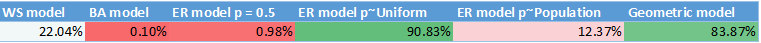}
\vspace{-.6cm}\caption{Coverage ratios, showing the average of 10 runs of the generators.\label{fig:bbox}
}
\end{figure}

We measure how representative a graph generator is by comparing pairwise correlations in the sample and in the ground truth. We measure how well a graph generator covers the range of values in the ground truth data by comparing the volumes of the generated data and the ground truth data. Specifically, for each generator we compare the volumes of the 10-dimensional bounding boxes for the ground truth set and the generated set. We consider a generator to be covering the ground truth set well if this ratio is close to $100\%$; see  Fig.~\ref{fig:bbox}.
\vspace{-.4cm}

\begin{figure}[t]
\includegraphics[width=\linewidth]{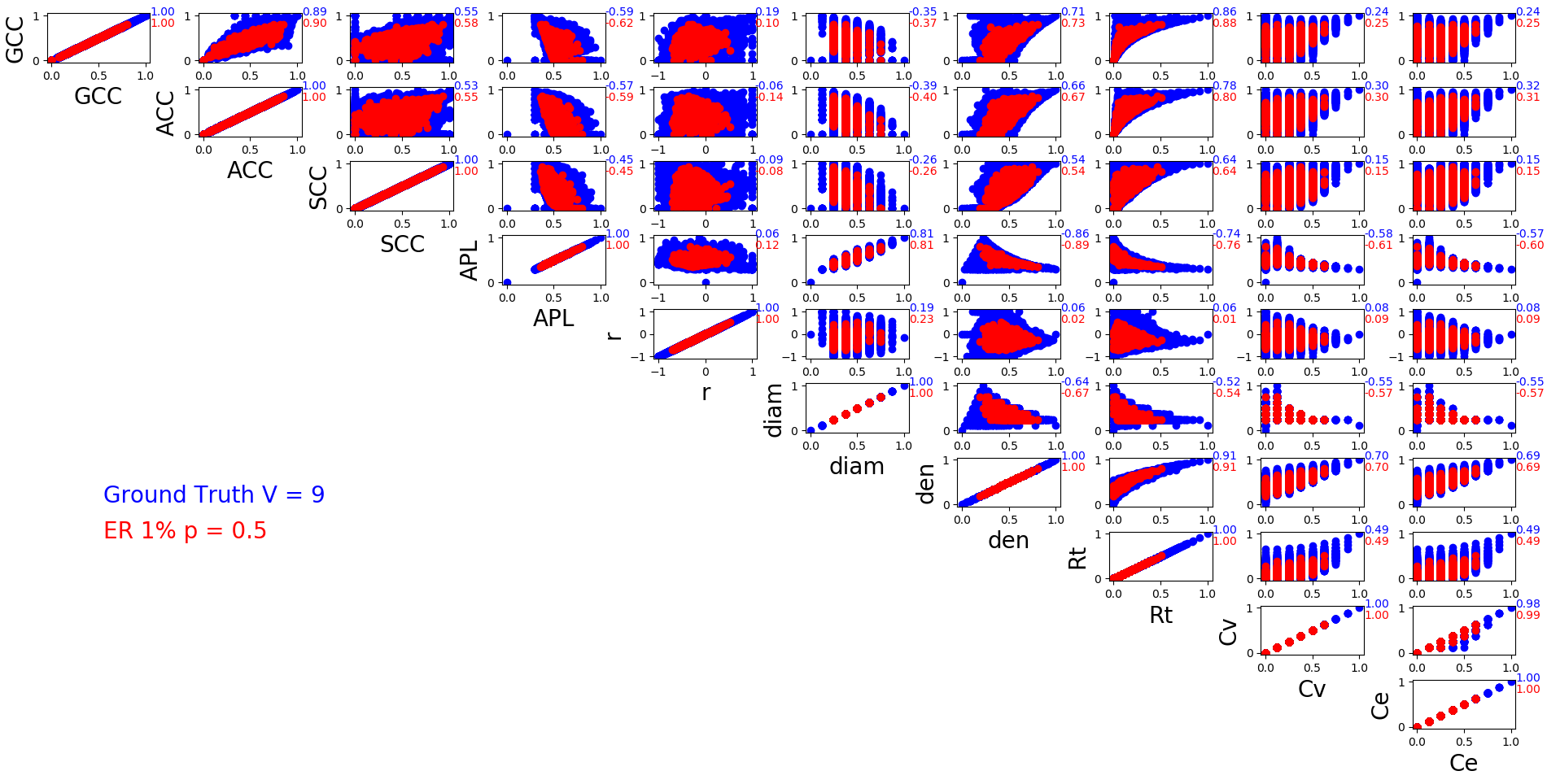}
\vspace{-.9cm}\caption{Ground truth (blue) and ER with $p=0.5$ (red).}
\label{fig:ER-half}
\end{figure}

Both of these measures can  be visualized by plotting each of the graphs in  ground-truth graph set as dots in the 2D matrix of correlations and then drawing the generated graph set on top of the first plot to see how well the generator set covers the ground-truth graph set. We color the ground-truth graph set in blue and the generated data in red. Because the ground-truth graph set includes all possible graphs for a fixed $|V|$,  there is at least one blue point under each red point. Detailed illustrations can be found in the full version of the paper~\cite{2018arXiv180809913C} but here we include one example of the most representative model: ER with $p=0.5$; see Fig.~\ref{fig:ER-half}. From this figure it is easy to see that nearly all pairwise correlations are very similar in the ground truth and in the generated data. Note, however, that from the same figure we can also see that this generator does not cover the range of possible values in the ground truth data well (e.g., in the columns corresponding to APL, r, diameter and density, the leftmost and rightmost points in the plots are blue).




\section{Finding Different Graphs with the Same Statistics}
\vspace{-.2cm} 
While our exploration of graph statistics, correlation, and generation revealed some challenges, it is still possible to explore the fundamental question of whether we can identify graphs that are similar across some statistics while being drastically different across others.
To find graphs that are identical over a number of graph statistics and yet are different, we  use the ground truth data for small non-isomorphic graphs. For larger graphs, we use the graph generators together with some filters.

\begin{figure}
\vspace{-.2cm}\includegraphics[width=\linewidth]{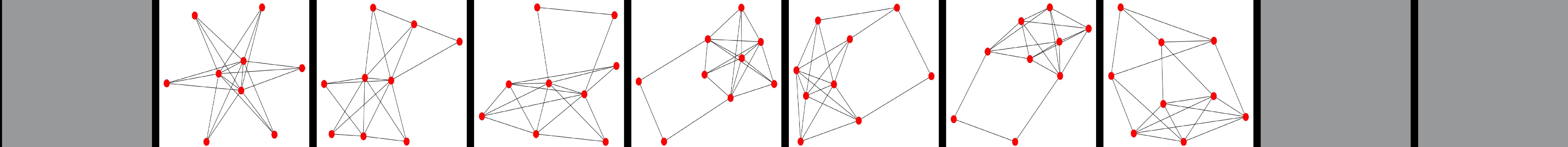}
\caption{Variability in assortativity.\label{fig:assort}}
\end{figure}

\smallskip\noindent{\bf Finding Graphs in the Ground Truth:} For $|V|\leq 10$, we directly use all possible non-isomorphic graphs as our dataset. 
In fact, we can fix different combinations of 5 statistics and still get multiple distinct graphs. We visualize this with figures that encapsulate the variability of one statistic in 10 slots, covering the ranges $[0.0, 0.1], [0.1, 0.2], \dots [0.9,1]$ and in each slot we show a graph (if it exists) drawn by a spring layout; see Fig.~\ref{fig:assort}.

For the first experiment, we fix $|V| = 9$, $APL\in (1.42,1.47)$, $den\in (0.52,0.57)$, $GCC\in$ (0.5,0.6), $R_t\in (0.15,0.25)$. 
Since all our statistics are normalized to $[0,1]$ and assortativity is in $[-1, 1]$, each of the ten slots has a range of 0.2. We find graphs for seven of the ten possible slots; see Fig.~\ref{fig:assort}. This figure also illustrates the output of our ``same stats, different graphs" generator: fix several statistics and generate graphs that vary in another statistic.

\begin{figure}
\vspace{-.2cm}\includegraphics[width=\linewidth]{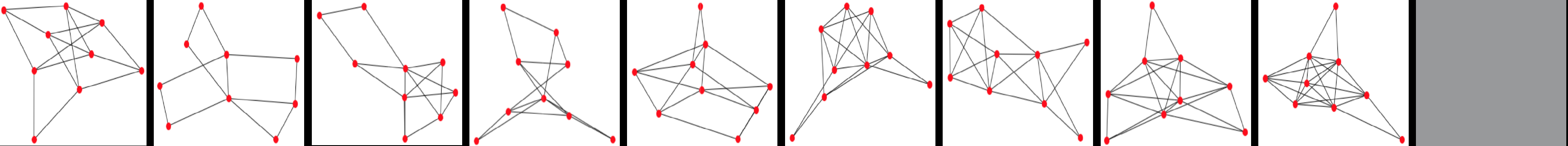}
\caption{Variability in GCC.\label{fig:gcc_vary}}
\end{figure}

Similarly, for the second experiment, we fix $|V| = 9$, $APL \in (1.47, 1.69)$, $diam = 3$, $Cv = 2$, $Ce = 2$, and $r \in (-0.22, -0.29)$ to obtain $GCC$ in the range $(0, 0.8)$; see Fig.~\ref{fig:gcc_vary}.

\begin{figure}[h]
\vspace{-.2cm}\hspace{.3cm}\includegraphics[width=.9\linewidth]{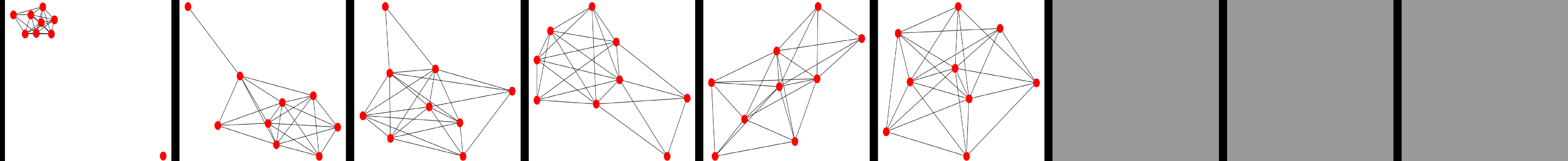}
\caption{Variability in edge connectivity.\label{fig:Ce}}
\end{figure}

As a final example, we fix $|V| = 9$, $SCC\in (0.75,0.85)$, $ACC\in (0.75,0.8)$, $r\in$ (-0.3, -0.2), $R_t\in (0.35,0.45)$ and find graphs with $C_e$ from 0 to 5; see Fig.~\ref{fig:Ce}.

Note that the graphs in Figures~\ref{fig:assort}-\ref{fig:Ce}. are different in structure even though they possess similar values for many properties. 

\smallskip\noindent{\bf Finding Graphs Using Graph Generators:} 
This approach relies on generating many graphs and filtering graphs based on several fixed statistics. 
For the two most important 
statistics of a graph, $|V|$ and $|E|$, we generate all graphs with a fixed $|V|$ and choose $|E|$ as follows:
\begin{enumerate}
\item uniform: select $|E|$ uniformly from its range. This is equivalent to forcing the edge density in the generated set to follow a uniform 
distribution;
\item population: select $|E|$ by forcing the edge density in the generated set to match the distribution in the ground truth (population) graph set.
\end{enumerate}

\begin{figure}[h]
\vspace{-.4cm}
\includegraphics[width=\linewidth]{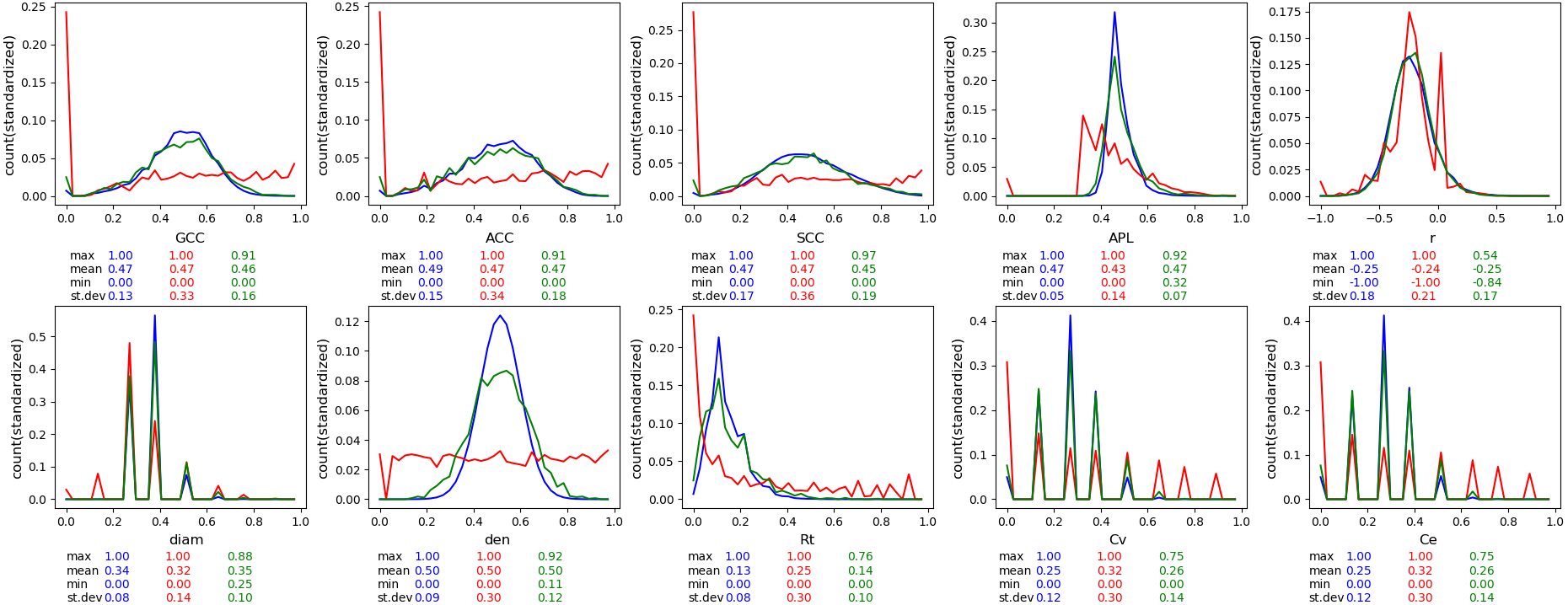}
\vspace{-.3cm}
\caption{Distribution of the ten statistics, including min/mean/max and standard deviation. Ground truth is in blue, population ER in green, uniform ER in red.\label{fig:sdd}}
\end{figure}

Using both edge selection strategies for all four generators, we compare the statistics distribution to the ground truth for $|V|=9.$ Figure~\ref{fig:sdd} illustrates how different statistics are distributed given uniform edge sampling and population-based edge sampling for the ER model. It shows that although the population-based sampling approach generates a distribution that is more similar to the ground truth, it has a narrower coverage (larger min and smaller max) than the uniform sampling. The WS and BA models also do not provide good coverage of the various statistics.

\vspace{-.5cm}
\section{Discussion and Future Work}
\vspace{-.2cm}

Random graph generators have been designed to model different types of graphs, but by design such algorithms sample the space of isomorphic graphs. For the purpose of studying graph properties and structure, we need generators that represent and cover the space of non-isomorphic graphs. 

We considered how to explore the space of graphs and graph statistics that make it possible to have multiple graphs that are identical in a number of graph statistics, yet are clearly different. To ``see" the difference, it often suffices to look at the drawings of the graphs. However, as graphs get larger, some graph drawing algorithms may not allow us to distinguish differences in statistics between two graphs purely from their drawings. 
We recently studied how the perception  statistics,  such as density and ACC, is affected by different graph drawing algorithms~\cite{soni2018perception}. The results confirm the intuition that some drawing algorithms are more appropriate than others in aiding viewers to perceive differences between underlying graph statistics. Further work in this direction might help ensure that differences between graphs are captured in the different drawings.


\bibliographystyle{splncs03}

\input{graph.bbl}
\end{document}

%% file: input/propertytable.tex
\begin{table}
\caption{The set of graph statistics considered in this paper.} 
\label{Table:properties}
{\renewcommand{\arraystretch}{1.5}%
\vspace{-.1cm}\begin{tabular}{ m{3cm} m{6.5cm} m{2cm} } 

\hline\hline 
  Name & Formula & Reference\\ 
\hline 

\multirow{2}{*}{\shortstack{Average Clustering \\Coefficient}}&  $ACC(G) = \frac{1}{n} \sum_{i=1}^n c(u_i), u_i \in V, n = |V| $ & \multirow{2}{*} {~\cite{chakrabarti2007graph,li2011graph,kairam2012graphprism,chakrabarti2006graph,mislove2007measurement}}
\\ & $c(v) = \frac{|\{(u,w)|u,w \in \Gamma(v) , (u,w) \in E\}|}{|\Gamma(v)|(|\Gamma(v)|-  1)/2}, v,u,w \in V$  &  \\

\shortstack{Global Clustering \\Coefficient} 
& $GCC(G) =\frac{ 3 \times |triangles|}{|connected\ triples|\ in\ the \ graph}$& ~\cite{chakrabarti2006graph,kairam2012graphprism}  \\

Square Clustering& $SCC(G) = \frac{\sum_{u=1}^{k_v}\sum_{w=u+1}^{k_v}q_v(u,w)}{\sum_{u=1}^{k_v}\sum_{w=u+1}^{k_v}[a_v(u,w)+q_v(u,w)] }$&~\cite{lind2005cycles}\\

Average Path Length& $APL = ave\{\frac{n-1}{ \sum_{v\in V} d(u,v), u\ne v}\}$&~\cite{chakrabarti2007graph,li2011graph,chakrabarti2006graph,mislove2007measurement}\\

Degree Assortativity& $r = \frac{\sum_{xy}xy(e_{xy}-a_xb_y) }{\sigma_a \sigma_b}$&~\cite{newman2003mixing,mislove2007measurement}\\

Diameter& $diam(G) = max\{dist(v,w), v,w \in V \}$ & ~\cite{chakrabarti2007graph,mcglohon2011statistical,kairam2012graphprism,mislove2007measurement}\\

Density & $den = \frac{2|E|}{|V|(|V| - 1)}$ & \\

Ratio of Triangles & Rt $= \frac{|triangles|}{|V|(|V| - 1)/2}$ & \\

Node Connectivity & Cv: the minimum number of nodes to remove to disconnect the graph & ~\cite{even1975network}\\

Edge Connectivity & Ce: the minimum number of edges to remove to disconnect the graph & ~\cite{even1975network}\\








\hline 
\end{tabular}}
\label{table:nonlin} 
\end{table}